\documentstyle[aps,preprint,pra]{revtex}
\begin{document}
\newcommand{\etal}{{\em et al.}\/}
\newcommand{\IP}{inner polarization}
\newcommand{\IPF}{\IP\ function}
\newcommand{\IPFs}{\IP\ functions}
\newcommand{\auth}[2]{#1 #2, }
\newcommand{\jcite}[4]{#1 {\bf #2}, #3 (#4)}
\newcommand{\et}{ and }
\newcommand{\twoauth}[4]{#1 #2 and #3 #4,}
\newcommand{\andauth}[2]{and #1 #2, }
\newcommand{\book}[4]{{\it #1} (#2, #3, #4)}
\newcommand{\erratum}[3]{\jcite{erratum}{#1}{#2}{#3}}
\newcommand{\JCP}[3]{\jcite{J. Chem. Phys.}{#1}{#2}{#3}}
\newcommand{\jms}[3]{\jcite{J. Mol. Spectrosc.}{#1}{#2}{#3}}
\newcommand{\jmstr}[3]{\jcite{J. Mol. Struct.}{#1}{#2}{#3}}
\newcommand{\cpl}[3]{\jcite{Chem. Phys. Lett.}{#1}{#2}{#3}}
\newcommand{\cp}[3]{\jcite{Chem. Phys.}{#1}{#2}{#3}}
\newcommand{\pr}[3]{\jcite{Phys. Rev.}{#1}{#2}{#3}}
\newcommand{\jpc}[3]{\jcite{J. Phys. Chem.}{#1}{#2}{#3}}
\newcommand{\jpcA}[3]{\jcite{J. Phys. Chem. A}{#1}{#2}{#3}}
\newcommand{\jpcB}[3]{\jcite{J. Phys. Chem. B}{#1}{#2}{#3}}
\newcommand{\jcc}[3]{\jcite{J. Comput. Chem.}{#1}{#2}{#3}}
\newcommand{\molphys}[3]{\jcite{Mol. Phys.}{#1}{#2}{#3}}
\newcommand{\cpc}[3]{\jcite{Comput. Phys. Commun.}{#1}{#2}{#3}}
\newcommand{\jcsfii}[3]{\jcite{J. Chem. Soc. Faraday Trans. II}{#1}{#2}{#3}}
\newcommand{\prsa}[3]{\jcite{Proc. Royal Soc. A}{#1}{#2}{#3}}
\newcommand{\jacs}[3]{\jcite{J. Am. Chem. Soc.}{#1}{#2}{#3}}
\newcommand{\ijqcs}[3]{\jcite{Int. J. Quantum Chem. Symp.}{#1}{#2}{#3}}
\newcommand{\ijqc}[3]{\jcite{Int. J. Quantum Chem.}{#1}{#2}{#3}}
\newcommand{\spa}[3]{\jcite{Spectrochim. Acta A}{#1}{#2}{#3}}
\newcommand{\tca}[3]{\jcite{Theor. Chem. Acc.}{#1}{#2}{#3}}
\newcommand{\tcaold}[3]{\jcite{Theor. Chim. Acta}{#1}{#2}{#3}}
\newcommand{\jpcrd}[3]{\jcite{J. Phys. Chem. Ref. Data}{#1}{#2}{#3}}

\draft
\title{
A fully {\em ab initio} quartic force field
of spectroscopic quality 
for SO$_3$\footnote{In memory of my colleague Dr. Jacqueline Libman 
OBM (1941--1997)}}
\author{Jan M.L. Martin}
\address{Department of Organic Chemistry,
Kimmelman Building, Room 262,
Weizmann Institute of Science,
IL-76100 Re\d{h}ovot, Israel. {\rm E-mail:} {\tt comartin@wicc.weizmann.ac.il}
}

\date{Submitted for special issue of {\em Spectrochimica Acta A} \today}
\maketitle
\begin{abstract}
The quartic force field of SO$_3$ was computed fully ab initio using
coupled cluster (CCSD(T)) methods and basis sets of up to $spdfgh$
quality. The effect of inner-shell correlation was taken into account.
The addition of tight $d$ functions is found to be essential for 
accurate geometries and harmonic frequencies. The equilibrium 
geometry and vibrational fundamentals are reproduced to within 0.0003 
\AA\ and (on average) 1.15 cm$^{-1}$, respectively. We recommend the
following revised values for the harmonic frequencies:
$\omega_1$=1082.7, $\omega_2$=502.6, $\omega_3$=1415.4, 
$\omega_4$=534.0 cm$^{-1}$.
In addition, we have shown that the addition of \IPFs\ to second-row 
elements is highly desirable even with more approximate methods like
B3LYP, and greatly improves the quality of computed geometries and
harmonic frequencies of second-row compounds at negligible extra 
computational cost. For larger such molecules, the B3LYP/VTZ+1 level 
of theory should be a very good compromise between accuracy and computational 
cost.
\end{abstract}

\section{Introduction}

The importance of the sulfuric anhydride (SO$_3$) molecule in 
atmospheric and industrial chemistry requires no further elaboration.

Experimental studies of its vibrational spectroscopy have to contend
with a number of problems such as the tendency of SO$_3$ to polymerize,
its very hygroscopic character, and its easy decomposition in the gas 
phase to SO$_2$. Possibly this explains why only a relatively small 
body of rotation-vibration spectroscopy data is available.

The only available anharmonic force field for this prototype planar 
XY$_3$ molecule is the work of Dorney, Hoy, and Mills\cite{Dor73} 
(DHM), who proposed two model force fields. Not enough data are 
available for an experimental force field refinement. 

In recent years, however, the methodology for computing anharmonic 
force fields fully ab initio has developed to the point where the 
computed force fields are at least \cite{h2co,n2o} suitable starting material for
further experimental refinement\cite{h2corefin,n2orefin,co2h2corefin} and at best
deliver essentially spectroscopic accuracy in their own 
right\cite{h2o,c2h2,fcch,c2h4more}. (By analogy with Boys' concept of 
`chemical accuracy', $\pm$1 kcal/mol, for energetics, we will 
arbitrarily define `spectroscopic accuracy' here as $\pm$1 cm$^{-1}$
on vibrational transition energies.)

Recently, the present author published a calibration study\cite{so2}
on the anharmonic force field of sulfur dioxide. It was found there
that the computed geometry and harmonic frequencies were critically
dependent on the presence of high-exponent `{\IPFs}' in the basis set,
whose contribution is actually much more important than that of 
inner-shell correlation. (They also contribute as much as 8 kcal/mol to
the total atomization energy of SO$_2$\cite{so2,Bau95}.
This is an extreme version of a phenomenon
that appears to occur more generally in second-row 
compounds\cite{sio,h2sio}.) Our best computed force field 
agreed to within 0.0004 \AA\ and 0.03 degrees with experiment for the
geometry: the 
errors in the fundamentals of SO$_2$ were +3.9, -0.4, and +0.4 cm$^{-1}$.
It would therefore appear that the same level of theory would be 
sufficient to produce a reliable force field for SO$_3$ as well. 

Some other recent ab initio anharmonic force field 
calculations on planar XY$_3$ systems include the work of Botschwina 
and coworkers\cite{BotsCF3+} on CF$_{3}$ and CF$_{3}^{+}$, of 
Schwenke\cite{Sch96} on CH$_3$, of Green et al.\cite{GreSiH3+} on 
SiH$_3^{+}$, of Martin and Lee\cite{bh3} on BH$_3$, and of 
Pak and Woods\cite{Pak97} on BF$_3$ and CF$_{3}^{+}$.

Previous ab initio studies of the vibrational force field were limited
to the harmonic part (e.g.\cite{Fla92}, which used scaled quantum 
mechanical (SQM)\cite{sqm} techniques as well as MCSCF calculations).
To the author's knowledge, this paper presents the first accurate
anharmonic force field for SO$_3$ obtained by any method. It will also
be shown that the harmonic frequencies derived from experiment by DHM
need to be substantially revised.

\section{Methods}

All electronic structure calculations were carried out using
MOLPRO 96.4\cite{molpro} with the \verb|tripu| and \verb|scfpr0|
patches\cite{tripu} installed, or a prerelease version of 
MOLPRO 97.3\cite{m97} running on a DEC Alpha 500/500 workstation
and an SGI Origin 2000 minisupercomputer at the Weizmann Institute
of Science.

As in our previous work on SO$_2$\cite{so2}, the CCSD(T) 
electron correlation method\cite{Rag89,Wat93}, as implemented 
by Hampel \etal\cite{Ham92}, has been used
throughout. The acronym stands for coupled cluster with all single
and double substitutions\cite{Pur82} augmented by a quasiperturbative 
account for triple excitations\cite{Rag89}. From extensive 
studies (see \cite{Lee95} for a review) this method is known to yield
correlation energies very close to the exact $n$-particle solution 
within the given basis set as long as the Hartree-Fock determinant is
a reasonably good zero-order reference wave function.
From the value of the ${\cal T}_1$ diagnostic\cite{Lee89}
calculated for SO$_3$, 0.018,
we see that this condition is definitely fulfilled.

Calculations including only valence correlation were carried out
using correlation consistent polarized $n$-tuple zeta (cc-pV$n$Z)
basis sets of 
Dunning and coworkers\cite{Dun89,Woo93a}. For 
sulfur/oxygen, the cc-pVTZ and cc-pVQZ basis sets correspond to
$(15s9p2d1f/10s5p2d1f)$ and $(16s11p3d2f1g/12s6p3d2f1g)$ 
primitive sets contracted to $[5s4p2d1f/4s3p2d1f]$ and
$[6s5p3d2f1g/5s4p3d2f1g]$, respectively.
Because of the strongly polar character of the SO bonds,
we have also considered the
aug-cc-pV$n$Z basis sets\cite{Ken92,Woo93b},
which consist of cc-pV$n$Z basis sets
with one low-exponent `anion' function added to each angular
momentum. 
In the interest of brevity, the standard acronyms cc-pV$n$Z
and aug$'$-cc-pV$n$Z will be replaced by V$n$Z and AV$n$Z,
respectively.

Furthermore, we considered the addition of a tight $d$ function
with exponent $\alpha_D$=3.203\cite{sio} to the sulfur basis set; its presence
is indicated by the notation VTZ+1, VQZ+1, and the like.

As in our previous studies on SO and SO$_2$ \cite{so2}, H$_2$SiO \cite{h2sio},
and various second-row diatomics\cite{sin,sio}, 
core correlation was included using the Martin-Taylor
basis set\cite{hf,cc}. This is generated by completely decontracting
the cc-pVTZ basis set and adding a single high-exponent $p$ function
and two even-tempered series, one of three tight $d$ functions and 
another of two tight $f$ functions.
The exponents are defined as $3^n\alpha$ (rounded to the nearest 
integer or half-integer), where $\alpha$ represents the
highest exponent of that angular momentum already present in the
cc-pVTZ basis. Obviously, such a basis set already amply covers the
`inner polarization' region as well. For brevity, CCSD(T)/Martin-Taylor
calculations with only the valence electrons correlated are 
denoted CCSD(T)/MTnocore, while in CCSD(T)/MTcore calculations,
all orbitals except for the sulfur $(1s)$-like orbital (which lies
too deep to interact appreciably with the valence shell) have been
correlated.

Geometry optimizations were carried out by repeated multivariate
parabolic interpolation with a step size of 0.001 bohr or radian, and 
a convergence threshold of about 10$^{-5}$ bohr or radian. Quartic
force fields were set up by finite differentiation
in symmetry-adapted coordinates
\begin{eqnarray}
S_1&=&(r_1+r_2+r_3)/\sqrt{3}\\
S_2&=&\delta\equiv \frac{\vec{r_1}.\vec{r_2}\times\vec{r_3}}{r_1r_2r_3}\\
S_{3a}&=&(2r_1-r_2-r_3)/\sqrt{6}\\
S_{3b}&=&(r_2-r_3)/\sqrt{2}\\
S_{4a}&=&(2\theta_1-\theta_2-\theta_3)/\sqrt{6}\\
S_{4b}&=&(\theta_2-\theta_3)/\sqrt{2}
\end{eqnarray}
in which the $r_i$ are the SO bond distances, the $\theta_i$ are the 
OSO bond angles opposite the corresponding $r_i$, and $\delta$ represents
the out-of-plane motion.
In order to keep higher-order contamination in the quartic portion of
the force field to a minimum, fairly small step
sizes 0.01 \AA\ and radian were used and CCSD(T) energies converged to
essentially machine precision.
Generation of the displaced Cartesian geometries and transformation
of the internal coordinate force field to Cartesian coordinates
were carried out with the help of the INTDER\cite{intder}
program. The anharmonic 
spectroscopic analysis was carried out by standard second-order
rovibrational perturbation theory using
a modified version of SPECTRO\cite{spectro,Gaw90}. No Fermi resonances
needed to be accounted for, but some of the rotation-vibration interaction
constants needed to be deperturbed for a Coriolis resonance $\omega_2\approx\omega_4$ around the $b$ axis.

\section{Results and discussion}

Computed and experimentally derived bond distances and harmonic frequencies
are given in Table \ref{tab:harm}, while computed and observed 
fundamentals can be found in Table \ref{tab:fund}. 

High-resolution values are available for all 
fundamentals except $\nu_1$, for which the 
Bondybey and English (BE)\cite{Bon85} 
value of 1068.6 cm$^{-1}$ appears to be
the most reliable one available. 
$\nu_3$=1391.5205 cm$^{-1}$ 
was taken from the work of Henfrey and Thrush (HT)\cite{Hen83}. 
The rotational fine structures of the
$\nu_2$ and $\nu_4$ bands overlap as well as exhibit Coriolis 
resonance: Kaldor et al. (KMDM) \cite{Kal73} resolved this spectrum to
give $\nu_2$=497.55 and $\nu_4$=530.18 cm$^{-1}$. They also 
definitively refuted a suggestion by Thomas and Thompson (TT)\cite{Tho70} 
that the assignments be reversed: the TT band origins (with the correct
assignment) are $\nu_2$=498.5 and $\nu_4$=529.16 cm$^{-1}$. 
A  high-resolution study by Ortigoso, Escribano, and Maki 
(OEM)\cite{Ort89} finally yielded $\nu_2$=497.5679(1) 
and $\nu_4$=530.0863(1) cm$^{-1}$. 
In the process, OEM also obtained improved values of the ground state
rotational constants, and particularly a revised $r_e$=1.41732 \AA,
which is near the lower limit of the earlier DHM value,
$r_e$=1.418$_4$$\pm$0.0010 \AA.

Comparing the observed with the calculated fundamentals, it is 
immediately seen that the CCSD(T)/VTZ fundamentals leave 
a lot to be desired: errors with respect to the experimental values
(BE,OEM,HT) are -25.1, -15.5, -20.4, and -16.1 cm$^{-1}$,
respectively, for $\nu_1$, $\nu_2$, $\nu_3$, and $\nu_4$. 
Switching to a VTZ+1 basis set (i.e. adding the tight $d$ function
on S) dramatically cuts these errors
to -1.6, +1.0, +4.8, and -2.0 cm$^{-1}$. 

Turning to Table \ref{tab:harm}, we see that in addition $r$(SO) is
shortened by no less than 0.0116 \AA, thus cutting 
the discrepancy with experiment
by more than half. Actually, the tight $d$ function appears to be even
more important for the quality of the results than increasing the 
basis set from VTZ to VQZ. Adding the \IPF\ to the VQZ basis set
has a smaller effect than for the VTZ basis set (as it well should,
since the VQZ basis set contains tighter $d$ functions than its VTZ
counterpart), but still affects the bond length by -0.0050 \AA.
The main difference between the VTZ+1 and VQZ+1 harmonics is that
$\omega_3$ drops by about 4.5 cm$^{-1}$, removing the largest error
remaining in the fundamentals at the CCSD(T)/VTZ+1 level.

Given the strongly polar character of the S-O bonds in SO$_3$, the
addition of anion functions is expected to have a nontrivial effect.
From VTZ+1 to AVTZ+1, the harmonics are lowered by no less
than -13.2, -5.7, 
-20.1, and -8.0 cm$^{-1}$, respectively; from VQZ+1 to AVQZ+1,
the effect is much weaker: -3.6, -2.4, -6.0, and -2.4 cm$^{-1}$. 
While the addition of diffuse functions lengthens $r$(SO) by +0.0030 
\AA\ between VTZ+1 and AVTZ+1, this likewise becomes much less 
significant from VQZ+1 to AVQZ+1 (+0.0009 \AA). 

Computing anharmonic corrections with these large basis sets for 
a four-heavy atom molecule is beyond the presently available 
computational resources. However, it is seen here that even from 
CCSD(T)/VTZ to CCSD(T)/VTZ+1, the basis set effect on the anharmonic corrections 
is very modest (-0.10, -0.27, -0.07, and +0.01 cm$^{-1}$), while for
SO$_2$\cite{so2}, their basis set dependence was likewise found to be
quite mild. We therefore opt for the CCSD(T)/VTZ+1 anharmonicities 
and will merely substitute an improved bond distance and improved 
harmonic frequencies in the analysis.

Doing the latter with the CCSD(T)/AVQZ+1 data yields harmonic 
frequencies which are systematically too low, with $r_e$ of course
still being 0.0053 \AA\ longer than experiment. 

A complete CCSD(T)/MTcore harmonic 
frequency calculation proved too demanding in terms
of computational resources: what could be succesfully completed was
a univariate optimization of $r_e$, with $\omega_1$ being obtained as a
by-product. 
The CCSD(T)/MTcore $r_e$ falls within the error bar of the DHM value
and is just 0.00032 \AA\ longer than the OEM value.
The contributions of inner-shell correlation to $r_e$ and $\omega_1$ 
turn out to be much more modest ($-$0.00283 \AA, +4.2 cm$^{-1}$) than 
those of inner polarization.

If we consider the ratios 
$\omega_i(\hbox{MTcore})/\omega_i(\hbox{MTnocore})$ for $\omega_1$
and for the harmonic frequencies of SO$_2$, we find that these ratios
are remarkably similar: 1.0039, 1.0037, 1.0038, and 1.0036. (They 
correspond almost exactly to 
$[r_e(\hbox{MTcore})/r_e(\hbox{MTnocore})]^{2}$.) Under these 
circumstances, we might be able to obtain `best estimate' harmonic
frequencies by scaling the CCSD(T)/AVQZ+1 values by
$\omega_1(\hbox{MTcore})/\omega_1(\hbox{MTnocore})=1.00386$,
which leads to the following values: 1081.2, 503.1, 1415.1, and 531.6
cm$^{-1}$. 

We now substitute these harmonic frequencies and the CCSD(T)/MTcore 
geometry in the spectroscopic analysis, and thus obtain the 
fundamentals labeled `best' in Table \ref{tab:fund}. Compared to
the (BE,OEM,HT) set of experimental values, the remaining discrepancies
are -1.5, +0.4, -0.3, and -2.4 cm$^{-1}$, or 1.15 cm$^{-1}$ on average.
We can thus safely claim `spectroscopic accuracy' for our best force field.

Turning now to the rotational constants (Table \ref{tab:rot}), we see 
that the computed and experimentally derived (OEM) $B_e$ agree to -0.02
\%: $C_e$ brings no additional information since it is fixed by the 
planarity relation $C_e=B_e/2$. From our computed rotational constants
we can however now determine computed $B_0$ and $C_0$, which likewise
turn out to be in excellent agreement with the observed values: -0.06 \%
and -0.03 \%, respectively. ($B_0$ and $C_0$ are independent data 
because of the inertial defect.) The fact that the rotational constants
are consistently computed slightly too small is consistent with our 
bond distance being slightly longer than the true value. 
From the relationship (in this case) $r_0=r_e\sqrt{B_e/B_0}$, we find 
$r_0$(calc.)=1.42004 \AA, compared to an experimental\cite{Kal73b}
value $r_0$(obs.)=1.4198$\pm$0.0002 \AA, and one derived from the OEM
$B_0$ of 1.41963 \AA. 
The discrepancy of +0.0004 \AA\ between the 
computed and OEM-derived $r_0$ is consistent with the discrepancy of
+0.0003 \AA\ between the computed and OEM $r_e$ values.
Aside from the observation that this would be considered 
excellent agreement between theory and experiment 
by any reasonable standard, this
suggests 
that the OEM equilibrium bond distance of 1.41732 \AA\ would be 
accurate to 0.0001 \AA\ or better. 
The computed $r_g$
(electron diffraction) distance from our force field, 1.42275 
\AA, is substantially longer than the experimental electron 
diffraction result\cite{Cla71}, 1.4188$\pm$0.003 \AA. Since this latter
study also finds an unrealistically short $r_e$=1.414$_2$ and 
$r_z$=1.416$_9$ \AA (our own computed $r_z$=1.42143 \AA), we can 
safely conclude that the electron diffraction result is in error.
(For a review of the different types of bond distances discussed,
see Ref.\cite{kuchitsu}.)

Our best harmonic frequencies agree relatively well with those obtained
by DHM from the experimental fundamentals and 
a valence model for anharmonicity, except for $\omega_1$ which still 
is about 16 cm$^{-1}$ too low in the better of their two models.
(The discrepancies for $\omega_3$ and $\omega_4$ are still 
substantially bigger than the differences between the computed and 
observed fundamentals, and it can safely be stated that the present ab
initio values are considerably more reliable than the experimentally
derived ones.) It thus comes as no surprise that the anharmonicity
constants (Table \ref{tab:anhar})
obtained by DHM from their valence models (particularly
$\omega_1-\nu_1$, which has the wrong sign) differ profoundly
from the presently computed set, which clearly is the more reliable 
one.

A set of `experimental' harmonic frequencies can be derived from our
best force field and the observed fundamentals by iteratively 
substituting $\omega_i^{[n+1]}=\omega_i^{[n]}+\nu_i^{\rm [expt.]}
-\nu_i^{[n]}$ in the spectroscopic analysis. The values thus obtained
are given as the entry labeled `Recommended' in Table \ref{tab:harm}.

Coriolis and rotation-vibration coupling constants can be found
in Table \ref{tab:anhar}.
The computed Coriolis coupling constant $B_e\zeta_{24}^B$ is found
as 0.1764 cm$^{-1}$ from our best force field, which agrees reasonably,
but not very, well with the OEM `Fit I' value of 0.191694(460) 
cm$^{-1}$. They note that their fit exhibits very strong dependency 
between $B_e\zeta_{24}^B$ and such  
parameters as the rotational $l$-doubling constant $q_4$
and the rovibrational coupling constants $\alpha_{2B}$ and 
$\alpha_{4B}$. Upon constraining $B_e\zeta_{24}^B$ to the 
force-field derived value of 0.1801, they found (Fit II in their paper) 
that all the abovementioned constants change drastically. It is 
noteworthy that our computed $\alpha_{2B}$=0.00031, 
$\alpha_{4B}$=-0.00050,
and $q_4$=0.00063 cm$^{-1}$ agree {\em much} better with the `Fit II' 
values of 0.000150, -0.000404, and 0.000497 cm$^{-1}$, respectively,
than with the
`Fit I' values of -0.000381, -0.000140, and -0.000047 cm$^{-1}$,
respectively. Linear extrapolation 
suggests that much of the residual discrepancies between our computed 
and the `Fit II' would disappear if  our $B_e\zeta_{24}^B$ value had 
been substituted in the experimental analysis.

OEM also obtained $\alpha_{2C}$ and $\alpha_{4C}$ values, 
which are in excellent agreement with our calculations, as are the
$\alpha_{3B}$ and $\alpha_{3C}$ values of HT.
Our computed $\alpha_{1B}$ and $\alpha_{1C}$ are larger
than the model-derived values of DHM, which appear to be on the low
side for all other constants as well.
The centrifugal distortion constants are quite small, and agree with
the experimental values of OEM to within the latter's uncertainties.

The trends in the computed quadratic force constants (Table \ref{tab:f2})
closely parallel those in the harmonic frequencies. (The most striking 
difference with the DHM quadratic force constants, which reproduce the
fundamentals rather than either of their sets of harmonic frequencies,
lies in the stretch-bend coupling constant $F_{34}$.)

In order to stimulate further research on the vibrational spectrum of
SO$_3$, the symmetry-unique cubic and quartic force constants in 
symmetry coordinates have been made available
in Table \ref{tab:ff}. The force fields in Cartesian, symmetry, and
normal coordinates can also be downloaded in 
machine-readable form on the World Wide Web at the Uniform Resource 
Locator (URL)
\verb|http://theochem.weizmann.ac.il/web/Papers/so3.html|

\section{A note on lower-level calculations}

Some readers might wonder how well less computationally intensive 
methods would do for the mechanical properties of SO$_3$ and the other
sulfur oxides, and whether the presence of the tight $d$ functions is
still relevant at that accuracy level.

In order to answer these questions, we have carried out geometry 
optimizations and harmonic frequency calculations for SO, SO$_2$, and
SO$_3$ using the popular B3LYP density functional 
method\cite{Bec93,Lee88} as implemented in GAUSSIAN 94\cite{g94}.
The VTZ, VTZ+1, and AVTZ+1 basis sets were considered, as was the
popular 6-31+G* basis set.
The results are summarized in Table \ref{tab:b3lyp}.

It is immediately seen that the 6-31+G* basis set 
systematically overestimates bond lengths by no less than 0.035 \AA,
and (largely as a result thereof)
underestimates stretching frequencies by as much as
80 cm$^{-1}$ and the SO$_3$ out-of-plane bending frequency
by about 50 cm$^{-1}$. These errors are substantially reduced
by using the VTZ basis set. However, at very small additional expense,
the addition of a tight $d$ function on S leads to quite respectable
agreement with experiment: 
residual errors for the B3LYP/VTZ+1 harmonic frequencies are
+5.7 cm$^{-1}$ in SO,
\{+0.4,+16.2,+6.5\} cm$^{-1}$ in SO$_2$, and
\{-5.7,-8.4,-10.5,-6.4\} cm$^{-1}$ in SO$_3$.
Trends in the effect of the tight $d$ 
function closely parallel those seen at the CCSD(T) level, which is
not surprising since it is essentially an SCF rather than a dynamical
correlation effect.\cite{so2,sio} 

At the B3LYP/VTZ+1 level, all bond lengths are now within 
+0.006--0.008 \AA\ of experiment, and the OSO angle is in excellent 
agreement with experiment. 
Overall, performance with the VTZ+1 basis set 
is as good as we can reasonably hope to get (e.g.\cite{dft})
at the B3LYP level. 

It is therefore clear that the addition of tight $d$ functions to the
basis set is eminently worthwhile even for less than `benchmark quality'
calculations on second-row compounds, as well as that B3LYP/VTZ+1 
would represent an excellent compromise between accuracy and 
computational cost for geometry and frequency calculations on larger
second-row systems.

\section{Conclusions}

The first-ever accurate anharmonic force field for SO$_3$
has been obtained fully {\em ab initio}. We have been able to establish
that:
\begin{itemize}
\item $r_e$ is reproduced to within +0.0003 \AA, and the fundamentals 
to within 1.15 cm$^{-1}$, on average;
\item like for SO$_2$ and (to a lesser extent) for second-row 
compounds in general, the addition of tight $d$ functions (`\IPFs')
to the basis set is essential for accurate results;
\item the following revised values are recommended values for the harmonic
frequencies:
$\omega_1$=1082.7, $\omega_2$=502.6, $\omega_3$=1415.4, 
$\omega_4$=534.0 cm$^{-1}$;
\item our computed rovibrational coupling, rotational $l$-doubling, 
and Coriolis coupling constants suggest a preference for the set of
constants in `Fit II' in Ref.\cite{Ort89} (OEM) over those in `Fit I'.
\end{itemize}
In addition, we have shown that the addition of \IPFs\ to second-row 
elements is highly desirable even with more approximate methods like
B3LYP, and greatly improves the quality of computed geometries and
harmonic frequencies of second-row compounds at negligible extra 
computational cost. For larger such molecules, the B3LYP/VTZ+1 level 
of theory should be a very good compromise between expense and
accuracy.

\acknowledgments

JM is a Yigal Allon Fellow,
an Honorary Research Associate (``Onderzoeksleider
in eremandaat'') of the
National Science Foundation of Belgium (NFWO/FNRS), 
and the incumbent of the Helen and Milton A. Kimmelman Career 
Development Chair.
The DEC Alpha workstation at the Weizmann Institute was purchased with
USAID (United States Agency for International Development) funds.

\begin{table}
\caption{Equilibrium bond distance (\AA) and 
harmonic frequencies (cm$^{-1}$) of SO$_3$\label{tab:harm}}
\begin{tabular}{lcdddd}
           &  $r_e$ & $\omega_1$($a_1'$) & $\omega_2$($a_2''$) & $\omega_3$($e'$) & $\omega_4$($e'$)\\
\hline
CCSD(T)/VTZ       & 1.43753 & 1057.7  & 487.5  & 1395.1  & 518.1  \\
CCSD(T)/VTZ+1     & 1.42594 & 1081.1  & 503.6  & 1420.2  & 532.0  \\
CCSD(T)/VQZ       & 1.42780 & 1071.1  & 496.2  & 1405.6  & 526.8  \\
CCSD(T)/VQZ+1     & 1.42279 & 1080.6  & 503.6  & 1415.7  & 532.0  \\
CCSD(T)/AVTZ      & 1.44038 & 1043.9  & 482.4  & 1374.1  & 509.7  \\
CCSD(T)/AVTZ+1    & 1.42890 & 1067.9  & 497.9  & 1400.1  & 524.0  \\
CCSD(T)/AVQZ+1    & 1.42372 & 1077.0  & 501.2  & 1409.7  & 529.6  \\
CCSD(T)/MTcore    & 1.41764 & 1092.5  & ---    & ---     & ---\\
CCSD(T)/MTnocore  & 1.42047 & 1088.3  & ---    & ---     & ---\\
Best estimate$^a$ & 1.41764 & 1081.2  & 503.1  & 1415.1  & 531.6  \\
Experiment     & 1.41732\cite{Ort89},1.418$_4$\cite{Dor73}\\
`VF' model (DHM)\cite{Dor73} && 1048.08  & 503.81  & 1408.96  & 538.64\\
`Extended' model (DHM)\cite{Dor73} &&  1064.89 & 505.97 & 1410.00 & 535.62\\
Recommended$^b$ &&  1082.7 & 502.6 & 1415.4 & 534.0 \\ 
\end{tabular}

(a) harmonics obtained by scaling CCSD(T)/AVQZ+1 values with
    ratio of 1.00386 between CCSD(T)/MTcore and CCSD(T)/MTnocore value for
    $\omega_1$ (see text)

(b) this work. Obtained by iteration of harmonics with best computed 
force field to exactly reproduce experimental fundamentals.
\end{table}

\begin{table}
\caption{Anharmonic corrections (cm$^{-1}$) and fundamentals
(cm$^{-1}$) for SO$_3$\label{tab:fund}}
\begin{tabular}{ldddd}
&$\omega_1-\nu_1$ & $\omega_2-\nu_2$ & $\omega_3-\nu_3$ & $\omega_4-\nu_4$ \\
\hline
CCSD(T)/VTZ              &14.182  &5.342  & 23.989  & 3.911\\
CCSD(T)/VTZ+1            &14.083  &5.067  & 23.921  & 3.920\\
Best estimate$^a$           &14.143  &5.069  & 23.930  & 3.921\\
`VF' model (DHM)       &$-$19.9 &6.3    & 19.1    & 8.5\\
`Extended' model(DHM)  & $-$3.1 &8.4    & 20.1    & 5.4\\
\hline
 &   $\nu_1$ &$\nu_2$ & $\nu_3$  &  $\nu_4$\\
\hline
VTZ       &            1043.5 & 482.1  &  1371.1   &  514.0\\
VTZ+1     &            1067.0 & 498.6  &  1396.3   &  528.1\\
Best calc.$^a$    &    1067.1 & 498.0  &  1391.2   &  527.7\\
Expt.     &            1068.6$^b$ & 497.5679(1)$^c$ & 1391.5205$^d$ & 530.0863(1)$^c$\\
\end{tabular}

(a) from substituting CCSD(T)/MTcore geometry and `best estimate'
harmonic frequencies in spectroscopic analysis of CCSD(T)/VTZ+1 force field

(b) Ref.\cite{Bon85} (BE)

(c) Ref.\cite{Ort89} (OEM)

(d) Ref.\cite{Hen83} (HT)

\end{table}

\begin{table}
\caption{Computed and observed bond distances (\AA), 
rotational constants (cm$^{-1}$), and centrifugal 
distortion constants (cm$^{-1}$) of SO$_3$\label{tab:rot}}
\begin{tabular}{lll}
        &   Best calc. & Expt.\\
\hline
$r_e$   &   1.41764 & 1.41732$^a$, 1.4184$\pm$0.0010$^b$, 1.414$_2$$^c$\\
$r_0$   &   1.42004 & 1.4198$\pm$0.0002$^d$, 1.41963$^e$\\ 
$r_g$   &   1.42275 & 1.4188$\pm$0.003$^c$\\
$r_z$   &   1.42143 & 1.416$_9$$^c$\\
$B_e$   &   0.34962 & 0.34968$^a$, 0.34923$^b$\\
$C_e$   &   0.17481 & $B_e/2$\\
$B_0$   &   0.34844 & 0.3485439$^a$, 0.34857$^b$\\
$C_0$   &   0.17393 & 0.173984$^a$, 0.17402$^b$\\
$10^7D_J$&     3.092 &   3.096(8)$^a$\\
$10^7D_{JK}$&-5.452 &  -5.47(2)$^a$ \\
$10^7D_K$&     2.543 &   2.55$^f$\\
\end{tabular}

(a) Ref.\cite{Ort89} (OEM)

(b) Ref.\cite{Dor73} (DHM)

(c) Ref.\cite{Cla71}

(d) Kaldor and Maki\cite{Kal73}

(e) From $B_0$ of OEM

(f) from planarity relation $D_K=-(2D_J+3D_{JK})/4$

\end{table}

\begin{table}
\caption{Anharmonicity constants, rotation-vibration 
coupling constants, and rotational $l$-doubling 
constants of SO$_3$\label{tab:anhar}. All values are in cm$^{-1}$.
Constants marked with an asterisk have been deperturbed for
Coriolis resonance}
\squeezetable
\begin{tabular}{lrrrl}
    & Best calc.& `VF' model &`Extended' model & Experiment\\
    & (this work)& DHM & DHM & OEM\\
\hline
$\alpha_{1B}$ &  0.00081  & 0.00067    &     0.00067  &       \\
$\alpha_{2B}$ &  0.00468  & 0.00402    &     0.00425  &       \\
$\alpha_{2B}^*$& 0.00031  &            &              & -0.000381$^a$,
0.000150$^b$\\
$\alpha_{3B}$ &  0.00111  & 0.00103    &     0.00103  & 0.001132(1)$^c$\\
$\alpha_{4B}$ & -0.00268  &-0.00271    &    -0.00269  &       \\
$\alpha_{4B}^*$&-0.00050  &            &              & -0.000140$^a$,
-0.000404$^b$\\
$\alpha_{1C}$ &  0.00041  & 0.00033    &     0.00033  &       \\
$\alpha_{2C}$ & -0.00013  &-0.00026    &    -0.00019  & -0.000130$^d$      \\
$\alpha_{3C}$ &  0.00058  & 0.00054    &     0.00054  &  0.0005999(3)$^c$     \\
$\alpha_{4C}$ &  0.00015  & 0.00005    &     0.00008  &  0.000157$^d$      \\
$q_3$ & -0.00012  & 0.00002    &     0.00002  &       \\
$q_4$ &  0.00500  & 0.00345    &     0.00345  &       \\
$q_4^*$ &  0.00063  &            &              & -0.000047$^a$, 0.000483$^b$\\
$B\zeta_{24}$ & 0.1764   &       &              & 0.1917(5)$^a$, 0.1801$^b$\\
$X_{11}$ &   -2.031  & -1.37      &      -1.38 \\
$X_{12}$ &    1.113  & 10.86      &        5.08  &       \\
$X_{13}$ &   -9.388  & -7.02      &       -6.94  &       \\
$X_{14}$ &   -1.249  & 24.27      &       10.27  &       \\
$X_{22}$ &   -0.748  & -4.24      &       -3.21  &       \\
$X_{23}$ &   -4.403  & -1.22      &       -2.36  &       \\
$X_{24}$ &    0.284  & -1.98      &       -2.18  &       \\
$X_{33}$ &   -5.434  & -4.66      &       -4.63  &       \\
$X_{34}$ &   -3.553  &  1.00      &       -1.60  &       \\
$X_{44}$ &   -0.011  & -6.20      &       -2.63  &       \\
$G_{33}$ &    2.828  &  2.46      &        2.47  &       \\
$G_{34}$ &   -0.279  & -0.20      &       -0.18  &       \\
$G_{44}$ &    0.150  &  6.25      &        2.08  &       \\
$R_{44}$ &    2.528  &  ---       &        ---   &       \\
\end{tabular}

(a) Ref.\cite{Ort89} (OEM), Fit I.

(b) OEM, Fit II ($B\zeta_{24}$ constrained to force field value 0.1801 cm$^{-1}$).

(c) Ref.\cite{Hen83} (HT)

(d) Ref.\cite{Ort89} (OEM)
\end{table}

\begin{table}
\caption{Quadratic force constants 
for SO$_3$ in symmetry-adapted internal
coordinates. Units are aJ, \AA, and radian, and
the restricted summation convention is used.\label{tab:f2}}
\begin{tabular}{lrrrrr}
                 & $F_{11}$& $F_{22}$& $F_{33}$& $F_{34}$ & $F_{44}$\\
\hline
DHM\cite{Dor73}$^a$ & 5.374   & 0.309   & 5.269   & -0.460   & 0.617  \\
CCSD(T)/VTZ      & 5.27167 & 0.30839 & 5.24224 & -0.33254 & 0.61456 \\
CCSD(T)/VTZ+1    & 5.50748 & 0.32393 & 5.44096 & -0.36341 & 0.63755 \\
CCSD(T)/AVTZ     & 5.13533 & 0.30320 & 5.08914 & -0.33047 & 0.59721 \\
CCSD(T)/AVTZ+1   & 5.37410 & 0.31787 & 5.29261 & -0.36336 & 0.62080 \\
CCSD(T)/VQZ      & 5.40597 & 0.31526 & 5.32635 & -0.35288 & 0.62686 \\
CCSD(T)/VQZ+1    & 5.50201 & 0.32249 & 5.40691 & -0.36436 & 0.63458 \\
CCSD(T)/AVQZ+1   & 5.46613 & 0.31990 & 5.36281 & -0.36459 & 0.62976 \\
CCSD(T)/MTcore   & 5.62424 \\
CCSD(T)/MTnocore & 5.58098 \\
\end{tabular}

(a) reproduce older values of fundamentals, {\em not} harmonic frequencies

\end{table}

\begin{table}
\caption{Computed cubic and quartic force constants 
for SO$_3$ in symmetry-adapted internal
coordinates. Units are aJ, \AA, and radian, and
the restricted summation convention is used.\label{tab:ff}}
\vspace*{12pt}
\squeezetable
\begin{tabular}{lrr}
   &CCSD(T)/VTZ & CCSD(T)/VTZ+1 \\
   \hline
$F_{111}$                                 &  -7.31818 & -7.62772\\
$F_{122}$                                 &  -0.58764 & -0.60070\\
$F_{13a3a}=F_{13b3b}$                     & -20.91209 & -21.57875\\
$F_{14a4a}=F_{14b4b}$                     &  -0.98244 & -1.01975\\
$F_{13a4a}=F_{13b4b}$                     &  1.02568  & 1.05182\\
$F_{3a3a3a}=-F_{3a3b3b}$/3                &  -4.84233 & -4.99590\\
$F_{3a3a4a}=-F_{3a3b4b}/2=-F_{3b3b4a}$    &  -0.29699 & -0.32604\\
$F_{3a4a4a}=-F_{3a4b4b}=-F_{3b4b4a}/2$    &   0.37460 & 0.38143\\
$F_{4a4a4a}=-F_{4a4b4b}$/3                &  -0.15162 & -0.15666\\
$F_{1111}$                                &  6.05065 & 6.12886\\
$F_{1122}$                                &  0.40348 & 0.39441\\
$F_{2222}$                                &  0.06145 & 0.06940\\
$F_{113a3a}=F_{113b3b}$                   & 33.41738 & 33.72571\\
$F_{113a4a}=F_{113b4b}$                   & -1.13882 & -1.10147\\
$F_{114a4a}=F_{114b4b}$                   &  0.68428 &  0.71380\\
$F_{223a3a}=F_{223b3b}$                   & -0.29300 & -0.30274\\
$F_{223a4a}=F_{223b4b}$                   & -0.34408 & -0.36252\\
$F_{224a4a}=F_{224b4b}$                   &  0.26088 &  0.27144\\
$F_{13a3a3a}=-F_{13a3b3b}/3$              & 15.54250 & 15.63856\\
$F_{13a3a4a}=-F_{13b3b4a}=-F_{13a3b4b}/2$ &  0.70144 &  0.78582\\
$F_{13a4a4a}=-F_{13a4b4b}=-F_{13b4b4a}/2$ & -0.39367 & -0.47240\\
$F_{14a4a4a}=-F_{14a4b4b}/3$              &  0.25712 &  0.26586\\
$F_{3a3a3a3a}=F_{3a3a3b3b}/2=F_{3b3b3b3b}$&  8.41884 & 8.50409\\
$F_{3a3a3a4a}=F_{3a3b3b4a}=F_{3a3a3b4b}=F_{3b3b3b4b}$ &  0.04720 &  0.08160\\
$F_{3a3a4a4a}=F_{3b3b4b4b}$               &  0.22378 &  0.21243\\
$F_{3a3a4b4b}=F_{3b3b4a4a}$               & -0.36425 & -0.40564\\
$F_{3a3b4a4b}=2(F_{3a3a4a4a}-F_{3a3a4b4b})$\\
$F_{3a4a4a4a}=F_{3b4b4a4a}=F_{3a4a4b4b}=F_{3b4b4b4b}$ & -0.27446 & -0.29029\\
$F_{4a4a4a4a}=F_{4a4a4b4b}/2=F_{4b4b4b4b}$&  0.12569 &  0.12956\\
\end{tabular}
\end{table}

\begin{table}
\caption{Basis set convergence for geometries (\AA, degrees)
and harmonic frequencies (cm$^{-1}$)
of SO$_n$ ($n$=1--3) at the B3LYP level\label{tab:b3lyp}}
\begin{tabular}{llrrrrl}
      &           & 6-31+G*& VTZ    & VTZ+1  & AVTZ+1 & Experiment\\
\hline
SO    & $r_e$     & 1.5157 & 1.4998 & 1.4891 & 1.4888 & 1.48108\cite{Cle94}\\
      & $\omega_e$& 1118.3 & 1146.6 & 1157.4 & 1156.2 & 1150.695(8)\cite{Cle94}\\
SO$_2$& $r_e$     & 1.4655 & 1.4504 & 1.4367 & 1.4380 & 1.43076(13)\cite{Sai69}\\
      & $\theta_e$& 118.68 & 118.31 & 119.25 & 119.16 & 119.33(1)\cite{Sai69}\\
      & $\omega_1$& 1133.0 & 1164.7 & 1184.1 & 1177.8 & 1167.91(4)\cite{Laf93}\\
      & $\omega_2$&  498.4 &  516.5 &  522.6 &  519.6 & 522.21(3)\cite{Laf93}\\
      & $\omega_3$& 1315.9 & 1352.1 & 1388.3 & 1376.2 & 1381.82(2)\cite{Laf93}\\
SO$_3$& $r_e$     & 1.4543 & 1.4389 & 1.4259 & 1.4270 & 1.41732\cite{Ort89}\\
      & $\omega_1$& 1018.1 & 1051.2 & 1077.0 & 1071.4 & 1082.7$^a$\\
      & $\omega_2$&  452.9 &  476.3 &  494.2 &  492.2 & 502.6$^a$\\
      & $\omega_3$& 1334.7 & 1375.3 & 1404.9 & 1394.1 & 1415.4$^a$\\
      & $\omega_4$&  492.5 &  512.2 &  527.6 &  523.8 & 534.0$^a$\\
\end{tabular}

(a) Recommended values (this work).

\end{table}

\begin{thebibliography}{99}
\bibitem{tripu} Without the \verb|tripu| patch, all open-shell CCSD(T) energies
in cases with
more than one doubly occupied orbital in any given irreducible
representation are in error. The \verb|scfpr0| patch is required to permit
sufficiently tight SCF convergence criteria to guarantee an accuracy of 12
decimal places in the CCSD and CCSD(T) energies.
\end{thebibliography}
\end{document}